\RequirePackage{fix-cm}
\documentclass{svjour3}                     
\smartqed  
\usepackage{amsmath}
\usepackage{graphicx}
\usepackage{slashed}
\usepackage{epstopdf}
\usepackage{hyperref}
\usepackage{pgfplots}

\begin{document}

\title{Neutrino flavor oscillations in a curved space-time}


\author{Luca Visinelli}


\institute{Luca Visinelli \at
	      Department of Physics and Astronomy\\
	      The University of Utah, 115 South 1400 East \#201\\
              Salt Lake City, Utah 84112-0830, USA\\
              \email{luca.visinelli@utah.edu}
}

\date{Received: \today / Accepted: XXXX}

\maketitle

\begin{abstract}
Using a WKB approximation of the Dirac equation in a curved spacetime, we obtain the expression for the phase shift between two neutrino mass eigenstates in a generic gravitational field. We apply this expression to two specific space-time geometries, namely the Kerr-Newman metric describing a rotating and charged black hole, and the Friedmann-Robertson-Walker metric.
\end{abstract}

\section{Introduction}

The idea that the flavor of a neutrino can change when traveling a macroscopic distance is quite old~\cite{pontecorvo}~\cite{mns}~\cite{bilenky}, and it is among the best examples showing us that we are in need of physics beyond the current Standard Model~\cite{minkowski}~\cite{mohapatra}, which does not embed neutrino masses. Despite this fact, a firm treatment for the probability of conversion between two neutrino flavors has now been set up, both in the vacuum~\cite{kayser}~\cite{giuntikimlee1}~\cite{giuntikimlee2}~\cite{giuntikimlee3}~\cite{lipkin}~\cite{grimus}~\cite{beuthe}~\cite{giunti}~\cite{giunti_review}~\cite{smirnov} and in matter~\cite{wolfenstein}~\cite{mikheyev}.

Neutrino oscillations in the presence of a gravitational field were pioneered by Stodolsky~\cite{stodolsky}, and it was later shown that the interaction of neutrinos with the gravitational field finds various applications both in cosmology and astrophysics, like the possible effects in the emissions occurring in type-II supernova events~\cite{ahluwalia96}~\cite{kojima96} or in the strong gravitational field of an active galactic nucleus~\cite{piriz96}.

The theory of neutrino flavor oscillations in a gravitational field has been further discussed in the Schwarzschild metric~\cite{ahluwalia97}~\cite{fornengo97}~\cite{cardall97}~\cite{ahluwalia98}~\cite{habib99}~\cite{fornengo99}~\cite{ahluwalia04}~\cite{crocker04}~\cite{maiwa04}~\cite{godunov09}, the Kerr metric~\cite{konno98}~\cite{wudka01}, the Lense-Thirring metric~\cite{lambiase05}, and the Hartle-Thorne metric~\cite{geralico12}. Also, neutrino spin oscillation in a gravitational field have been discussed in Refs.~\cite{dvornikov02}~\cite{dvornikov05}~\cite{dvornikov06}~\cite{dvornikov13}~\cite{alavi13}.

In this paper we derive the phase shifts for free-streaming neutrinos using a WKB approximation~\cite{anandan} of the solution to the Dirac equation in curved space-time. We apply our results to describe neutrino flavor oscillations in two well-known space-time metrics, namely the Kerr-Newman metric, which describes the spacetime around a charged and rotating black hole, and the Friedmann-Robertson-Walker (FRW) metric, which describes the cosmological expansion of a homogeneous and isotropic Universe. Notice that the electric charge of a Kerr-Newman black hole affects the trajectory of the neutrino, despite such particle being free of electric charge. This occurs because the energy density associated with the electromagnetic field around the black hole enters the equation of general relativity and influences space-time around the black hole itself.

To the best of our knowledge, the oscillation of neutrinos in the two metrics presented has not been explored before. After discussing the Dirac equation for a fermion propagating in a generic pseudo-Riemann manifold $\mathcal{M}$ in Sec.~\ref{sec_dirac_curved}, we apply this formalism to neutrino oscillations in Sec.~\ref{neutrinos}. We derive the expression for the probability of neutrino oscillations in the Kerr-Newman metric in Sec.~\ref{sec_kerr}, and in the FRW metric in Sec.~\ref{sec_FRW}.

\section{Approximation of the Dirac equation in curved space-time} \label{sec_dirac_curved}

\subsection{Review of Dirac equation on a curved space-time}

Before discussing the properties of neutrinos in the presence of gravity, we briefly review the basic tools for the Dirac equation on curved space-time. On a flat, Minkowski background, the Dirac equation for a spinor field $\psi(x)$ of mass $m$ reads
\begin{equation} \label{dirac}
i\,\gamma^a\,\partial_a\,\psi(x) = m\,\psi(x).
\end{equation}
Here, a Latin index $a \in\{0,1,2,3\}$ labels the coordinates in the Minkowski space-time, while the Dirac matrices $\gamma^a$ satisfy the Clifford algebra
\begin{equation}\label{clifford_flat}
\{\gamma^a,\gamma^b\} = 2\eta^{ab},
\end{equation}
where curly brackets indicate the anti-commutation operation. In the following, the convention for the flat metric is $\eta_{ab} = {\rm diag}(1,-1,-1,-1)$.

According to general relativity, gravitational effects are included by expressing the Dirac Eq.~(\ref{dirac}) on a suitable manifold. We consider a four-dimensional, torsion-free pseudo-Riemann manifold $\mathcal{M}$ equipped with a metric tensor $g_{\mu\nu}$, where Greek indices are used to refer to coordinates on $\mathcal{M}$. Indices for quantities on $\mathcal{M}$ are lowered with $g_{\mu\nu}$ and raised with the inverse of the metric tensor $g^{\mu\nu}$, which is defined by
\begin{equation}
g^{\mu\lambda}\,g_{\lambda\nu} = \delta_\mu^\nu.
\end{equation}
The metric tensor defines the torsion-free Levi-Civita connection on $\mathcal{M}$ as
\begin{equation}
\Gamma_{\mu\nu}^\sigma = \frac{1}{2}g^{\sigma\lambda}\left(\partial_\nu\,g_{\lambda\mu}+\partial_\mu\,g_{\nu\lambda}-\partial_\lambda\,g_{\mu\nu}\right),
\end{equation}
which appears in the parallel transport of vector quantities. For spinors, we define the spinor connection on $\mathcal{M}$ as
\begin{equation} \label{spinor_connection}
\Gamma_{\mu} = \frac{1}{8}\omega_{ab\mu}\left[\gamma^a,\gamma^b\right],
\end{equation}
where squared brackets indicate the commutation operation. We defined the gravity spin connection,
\begin{equation}
\omega_{ab\mu} = e_a^\nu\,\partial_\mu\,e_{b\nu} - e_a^\nu\,\Gamma^\sigma_{\mu\nu}\,e_{b\sigma},
\end{equation}
in terms of the tetrad $e_{a}^{\mu}$ satisfying
\begin{equation}
g_{\mu\nu}\,e^\mu_a \,e^\nu_b = \eta_{ab}.
\end{equation}
Finally, we define the covariant derivative for a spinor field as
\begin{equation}
\mathcal{D}_\mu \equiv \partial_\mu + \Gamma_\mu.
\end{equation}
Given the Dirac equation on the flat space-time, the recipe for expressing the equation on the manifold $\mathcal{M}$ consists in replacing all derivative operators with the covariant derivative,
\begin{equation}
\partial_a \to e_a^\mu\,\mathcal{D}_\mu,
\end{equation}
so that Eq.~(\ref{dirac}) with this prescription reads
\begin{equation}
i\,e^\mu_a\,\gamma^a\,\mathcal{D}_{\mu}\,\psi(x) = m\,\psi(x).
\end{equation}
Defining the Dirac matrices on the manifold $\gamma^\mu \equiv e^\mu_a\,\gamma^a$ and satisfying the algebra
\begin{equation}
\{\gamma^{\mu},\gamma^{\nu}\} = 2g^{\mu\nu},
\end{equation}
we obtain the Dirac equation for the massive spinor field on $\mathcal{M}$,
\begin{equation} \label{dirac_curved}
i\,\gamma^\mu\,\mathcal{D}_\mu\,\psi(x) = m\,\psi(x).
\end{equation}
The action associated to the Dirac Eq.~(\ref{dirac_curved}) is
\begin{equation}
S = \int d^4x\,\sqrt{g}\,\mathcal{L}_D,
\end{equation}
where $g = g^{\mu\nu}\,g_{\mu\nu}$, and we introduced the Lagrangian
\begin{equation}
\mathcal{L}_D = \frac{i}{2}\,\left[\bar{\psi}\,\gamma^\mu\,\mathcal{D}_\mu\,\psi - (\mathcal{D}_\mu\,\bar{\psi})\,\gamma^\mu\,\psi\right] - m\,\bar{\psi}\,\psi.
\end{equation}

\subsection{WKB approximation}

We seek for an approximate solution to the Dirac Eq.~(\ref{dirac_curved}) using a Wentzel-Kramers-Brillouin (WKB) approximation. In the literature, various forms of the WKB approximations have been proposed. Stodolsky~\cite{stodolsky} decomposes the complex spinor $\psi(x)$ into an amplitude $\chi = \chi(x)$ and a semi-classical phase $S = S(x)$, as
\begin{equation} \label{ansatz_stodolsky}
\psi(x) = e^{-iS(x)/\hbar}\,\chi(x).
\end{equation}
Anandan~\cite{anandan} and Maiwa and Naka~\cite{maiwa04} further decompose the complex spinor $\psi(x)$ into a power series in terms of $\hbar$, as
\begin{equation} \label{ansatz_anandan}
\psi(x) = e^{-iS(x)/\hbar}\,\sum_{k=0}^{+\infty}\, \left(\frac{\hbar}{i}\right)^k\,\chi_k(x).
\end{equation}
Here, we include an additional phase containing the spin connection $\Gamma_\mu$, as
\begin{equation} \label{ansatz}
\psi(x) = e^{-iS(x)/\hbar}\,e^{-\Gamma_\mu\,x^\mu}\,\sum_{k=0}^{+\infty}\, \left(\frac{\hbar}{i}\right)^k\,\chi_k(x).
\end{equation}
Substituting Eq.~\eqref{ansatz} into Eq.~(\ref{dirac_curved}) and equating terms with the same power of $\hbar$, a system of recurring equations for the amplitudes $\chi_k$ is obtained as
\begin{equation} \label{ansatz1.1}
\left(\slashed{\partial}\,S - m \right)\,\chi_0 = 0,\quad \hbox{and}\quad \left(\slashed{\partial}\,S - m \right)\,\chi_k = \slashed{\partial}\,\chi_{k-1},\quad\hbox{for $k \neq 0$}.
\end{equation}
Notice that, since $S(x)$ is a scalar function, the covariant derivative of this function has been replaced with the ordinary derivative. Since the spin connection can be written in terms of the parity-violating matrix $\gamma^5$~\cite{cardall97},
\begin{equation} \label{spin_connection_gamma5}
\Gamma_\mu = (-g)^{-1/2}\,\frac{\gamma_5}{2i}\,A_\mu,\quad \hbox{with} \quad A^\mu = \frac{(-g)^{1/2}}{4}\,\epsilon^{abcd}\,e_a^\mu\,\left(e_{b\nu,\sigma}-e_{b\sigma,\nu}\right)\,e_c^\nu\,e_d^\sigma,
\end{equation}
we conclude that the extra phase factor in Eq.~\eqref{ansatz} explicitly accounts for the interaction between the metric and the spin orientation of the spinor (see Refs.~\cite{ahluwalia96}~\cite{ahluwalia97} for a discussion on the effects of this interaction).

Multiplying the expression for $\chi_0$ on the right by $\gamma^\nu\,\partial_\nu\,S(x) - m$, we obtain the Hamilton-Jacobi (HJ) equation
\begin{equation} \label{HJ}
g^{\mu\nu}\,\partial_{\mu}\,S(x)\,\partial_{\nu}\,S(x) = m^2.
\end{equation}
Eq.~(\ref{HJ}) expresses the mass-shell condition for a massive particle in a curved space-time, and allows us to identify the phase $S(x)$ with the classical action for a particle of mass $m$ moving on $\mathcal{M}$, provided that the four-momentum of the particle is~\cite{stodolsky}
\begin{equation} \label{4momentum}
p_{\mu} = m \,g_{\mu\nu}\,\frac{dx^{\nu}}{d\tau},
\end{equation}
and where the proper time $\tau$ is given by
\begin{equation}
d\tau^2 = g_{\mu\nu}\,dx^\mu\,dx^\nu.
\end{equation}
Eq.~(\ref{HJ}) is equivalent to the mass-shell condition if we identify
\begin{equation}
p_\mu = \partial_\mu\,S(x),
\end{equation}
which is the Hamilton-Jacobi relation. A solution to Eq.~(\ref{HJ}) is then~\cite{stodolsky}
\begin{equation} \label{action}
S(x) = \int^x \,p_{\mu}\,dx^{\mu},
\end{equation}
with the Lagrangian describing the geodesic being
\begin{equation} \label{lagrangian_geodesic}
\mathcal{L} = \sqrt{g_{\mu\nu}\,\frac{dx^\mu}{d\tau}\,\frac{dx^\nu}{d\tau}}.
\end{equation}
In fact, the Euler-Lagrange equation applied to the Lagrangian in Eq.~(\ref{lagrangian_geodesic}) gives the geodesic equation
\begin{equation} \label{geodesics}
\frac{d}{d\tau}\left(g_{\mu\nu}\,\frac{dx^\nu}{d\tau}\right) - \frac{1}{2}\,g_{\alpha\beta,\mu}\,\frac{dx^\alpha}{d\tau}\,\frac{dx^\beta}{d\tau} = 0.
\end{equation}

\section{Neutrinos} \label{neutrinos}

\subsection{Probability of oscillation} \label{Probability of oscillation}
We now apply the general results presented in Sec.~\ref{sec_dirac_curved} to the theory of neutrino oscillations. For the $i$-th neutrino generation with mass eigenvalue $m_i$, we consider a neutrino mass eigenstate $\nu_i(x)$ propagating from the point $x_A$, where the source is placed, to the receiver at $x_B$. Neutrinos produced at the source can be described by a flavor state $\nu_{\alpha}(x)$ ($\alpha = e, \mu, \tau$) that is a linear combination of the mass eigenstates $\nu_i(x)$ as
\begin{equation}
\nu_{\alpha}(x) = \sum_i\,U_{\alpha i}^*\,\nu_i(x).
\end{equation}
The mixing matrix $U$, also known as the Maki-Nakagawa-Sakata-Pontecorvo (MNSP) matrix \cite{mns}, is the leptonic analogous of the Cabibbo-Kobayashi-Maskawa matrix that governs quarks mixing. For three generations of neutrinos, the MNSP matrix is parametrized by three mixing angles $\theta_i$, a phase $\delta$ describing CP-violation and two additional phases $\alpha_1$ and $\alpha_2$ that may differ from zero only if neutrinos are Majorana particles, while $\alpha_1=\alpha_2 = 0$ if neutrinos are Dirac particles.

The amplitude for the process in which a neutrino of flavor $\alpha$ at position $x_A$ is detected as a neutrino of flavor $\beta$ at position $x_B$ is given by
\begin{equation} \label{flat1}
\mathcal{A}_{\beta\alpha} = \langle\nu_{\beta}(x_B)|\nu_\alpha(x_A)\rangle = \sum_i  \,U_{\alpha i}^*U_{\beta i}\,\langle\nu_i(x_B)|\nu_i(x_A)\rangle.
\end{equation}
We approximate the expression for the spinor $\nu_i(x)$ with the WKB in Eq.~\eqref{ansatz_stodolsky}, where both the action $S_i(x)$ for the $i$-th mass eigenstate and the spin connection $\Gamma_\mu$ appear. As discussed in Refs.~\cite{ahluwalia96}~\cite{ahluwalia97}~\cite{ahluwalia98}~\cite{ahluwalia04}, when discussing neutrino oscillation we might distinguish at least three different scenarios where neutrino oscillate in I) a flat spacetime, II) a curved spacetime in a non-rotating frame and III) a curved spacetime in a rotating frame. In Scenarios I) and II), the neutrino phase difference depends on $S(x)$ only, and not on the spin connection, so that the phase difference reads
\begin{equation} \label{action_stodolsky}
S(m_i,x_B - x_A) \equiv S_i(x_B) - S_i(x_A) = \int_{x_A}^{x_B} p_{\mu}\,dx^{\mu}.
\end{equation}
Since this phase difference is a scalar quantity, it is invariant in all frames.

In Scenario III), an extra contribution to the phase shift appears if two mixing eigenstates $\nu_i(x)$ have different spin orientation, as can be seen from the representation of the spin connection in Eq.~\eqref{spin_connection_gamma5}, where the parity-violating matrix $\gamma^5$ appears~\cite{cardall97}. This additional contribution might be seen as a gravitational analogue to the Zeeman effect~\cite{ahluwalia97}. In the following we will only consider the same spin orientation for the massive neutrinos and we will not treat this contribution. Since in this work we focus on neutrino flavor oscillations induced by the action $S(x)$, in the following we will consider the same spin orientation for the massive neutrinos only, neglecting the interaction of the neutrino spin with the metric resulting from $\Gamma_\mu$ (see Refs.~\cite{konno98}~\cite{dvornikov06}~\cite{alavi13} for additional discussion).

The expression for the probability amplitude of neutrino oscillations on a generic manifold is then
\begin{equation}
\mathcal{A}_{\beta\alpha} = \sum_i  \,U_{\alpha i}^*U_{\beta i}\,e^{-i S(m_i, x_B-x_A)}.
\end{equation}
Indicating the phase difference between two mass eigenstates with
\begin{equation} \label{relative_phase}
\Phi_{ij} = S(m_i, x_B-x_A) - S(m_j, x_B-x_A),
\end{equation}
the probability of conversion from flavor $\alpha$ to $\beta$ is
\begin{equation} \label{probability}
\mathcal{P}_{\beta\alpha} = \left|\mathcal{A}_{\beta\alpha}\right|^2 = \sum_{ij}  \,U_{\alpha i}^*U_{\beta i}U_{\alpha j}U_{\beta j}^*\,e^{-i \Phi_{ij}}.
\end{equation}
Eq.~(\ref{relative_phase}) can be specialized to the case of relativistic neutrinos, which is the case for various realistic situations. After all, the most stringent upper bound on the sum of the masses of the three known neutrinos species comes from cosmological consideration as $\sum_i m_i < 0.28$eV \cite{bernardis}, while neutrinos often take part in processes involving energies ranging from the keV to hundreds of GeVs.

Motivated by these considerations, we expand the neutrino action in powers of $m_i^2$ as
\begin{equation} \label{taylor_expansion}
S(m_i, x_B - x_A) = \sum_{k=0}^{+\infty} \,\frac{(m_i^2)^k}{k!}S^{(k)}(x_B-x_A),
\end{equation}
where
\begin{equation} \label{taylor_expansion}
S^{(k)}(x_B-x_A) = \frac{\partial^{(k)}\, S(m_i, x_B - x_A)}{\partial^{(k)}\,
m_i^2}\bigg|_{m_i^2=0}.
\end{equation}
For a similar treatment, see also Ref.~\cite{wudka01}. The phase difference in Eq.~(\ref{relative_phase}) reads
\begin{equation}\label{relative_phase_expanded}
\Phi_{ij} = \Delta m_{ij}^2 \,S^{(1)}(x_B - x_A) + \frac{1}{2}\,\Delta m_{ij}^4 \,S^{(2)}(x_B - x_A) + ...\,,
\end{equation}
with
\begin{equation}
\Delta m_{ij}^2 = m_i^2 - m_j^2,\quad \hbox{and}\quad \Delta m_{ij}^4 = m_i^4 - m_j^4 = \left(m_i^2 + m_j^2\right)\,\Delta m_{ij}^2.
\end{equation}
The term corresponding to $k=0$ in the Taylor expansion of Eq.~\eqref{taylor_expansion} is canceled in the difference in Eq.~\eqref{relative_phase} which defines the phase shift $\Phi_{ij}$.

\subsection{Two generations of neutrinos}

The result in Sec.~\ref{Probability of oscillation} can be applied to any number of neutrino generations and for any neutrino energy. However, in many cases of interest like the solar or atmospheric neutrino mixing, the probability of conversion to a specific neutrino flavor is suppressed. If this is the case the MNSP matrix reduces to an element of the SO(2) group and can be parametrized by one mixing angle only $\theta$ as
\begin{equation} \label{rotation_EM}
U =
\left( \begin{array}{cc}
\cos\theta & \sin\theta\\
-\sin\theta & \cos\theta\\
\end{array} \right).
\end{equation}
In the case of two effective neutrino generations, there is only one splitting in mass $\Delta m^2 = m_1^2-m_2^2$ (we assume $m_1 > m_2$), and $\Delta m^4 =  m_1^4-m_2^4$. We thus suppress the lower indices in $\Phi = \Phi_{12}$. The probability of conversion is
\begin{equation}\label{probability_conversion}
\mathcal{P}_{\beta\alpha} = \begin{cases}
\sin^22\theta\,\sin^2\Phi,\quad\quad\quad \beta \neq \alpha,\\
1-\sin^22\theta\,\sin^2\Phi,\quad\beta = \alpha.
\end{cases}
\end{equation}
With the series expansion given in Eq.~(\ref{relative_phase_expanded}), Eq.~(\ref{probability_conversion}) reads
\begin{equation}\label{probability_conversion_approx}
\mathcal{P}_{\beta\alpha} = \begin{cases}
\sin^2\theta\,\sin^2 \left[S^{(1)}(x_B - x_A)\,\Delta m^2 + \frac{1}{2}\,S^{(2)}(x_B - x_A)\,\Delta m^4\right],\quad\quad \beta \neq \alpha,\\
1 - \sin^2\theta\,\sin^2 \left[S^{(1)}(x_B - x_A)\,\Delta m^2 + \frac{1}{2}\,S^{(2)}(x_B - x_A)\,\Delta m^4 \right], \quad \beta = \alpha.
\end{cases}
\end{equation}
This is the general expression for the probability of flavor conversion of one neutrino propagating from the source to the receiver. In the next section, we apply the procedure described to obtain the phase difference in some well-known metric spaces.

\section{Application to the Kerr-Newman metric} \label{sec_kerr}

\subsection{General treatment}

Space-time around a charged, rotating body of mass $M$, angular momentum $J$, and charge $Q$, is described by the Kerr-Newman metric,
\begin{equation} \label{kerr_metric}
ds^2 = \left(1-\frac{\Lambda}{\rho^2}\right)\,dt^2 + \frac{2\Lambda\,a\,\sin^2\theta}{\rho^2}\,dt\,d\phi- \frac{\rho^2}{\Delta}\,dr^2 - \rho^2\,d\theta^2 - \left[a^2+r^2 + \frac{a^2\,\Lambda\,\sin^2\theta}{\rho^2}\right]\,\sin^2\theta d\phi^2,
\end{equation}
where, in Planck units, we defined the lengths $r_s = 2M$, $a = J/M$, and $r_Q = Q^2$. The Kerr-Newman metric reduces to the Kerr metric for $r_Q = 0$, to the Reissner-Nordstrom metric for $a = 0$, and to the Schwarzschild metric for $a = r_Q = 0$. We also introduced $\Lambda = r_s\,r - r_Q^2$, $\Delta = r^2 + a^2 - \Lambda$, and $\rho^2 = r^2 + a^2\,\cos^2\theta$. The inverse metric tensor is
\begin{equation} \label{inverse_kerr}
g^{\mu\nu} =
\left( \begin{array}{cccc}
g^{tt} & 0 & 0 & g^{t\phi}\\
0 & -\frac{\Delta}{\rho^2} &0 & 0\\
0 & 0 & -\frac{1}{\rho^2} & 0\\
g^{t\phi} & 0 & 0 & g^{\phi\phi}\\
\end{array} \right),
\end{equation}
where
$$g^{tt} = \frac{\rho^2\,\left(a^2+r^2\right)+ a^2\,\Lambda\,\sin^2\theta}{\Delta\,\rho^2} = \frac{(r^2+a^2)^2}{\Delta\,\rho^2} - \frac{a^2\,\sin^2\theta}{\rho^2},$$
$$g^{t\phi} = \frac{\Lambda\,a}{\Delta\,\rho^2},$$
$$g^{\phi\phi} = -\frac{\Delta - a^2\,\sin^2\theta}{\Delta\,\rho^2\,\sin^2\theta}.$$
Being stationary and axisymmetric, the Kerr metric admits two Killing vector fields $\partial_t$ and $\partial_\phi$, thus, both the energy $E$ and the azymuthal angular momentum $L$ of a particle are conserved. For this reason, the action can be separated as
\begin{equation}
S = E\,t - L\,\phi - W_r(r) - W_\theta(\theta),
\end{equation}
and the HJ Eq.~(\ref{HJ}) for massive neutrinos reads
\begin{equation}
g^{tt} \,E^2 - 2\,g^{t\phi}\,E\,L+g^{\phi\phi}\,L^2 + g^{rr}\,\left(\frac{dW_r}{dr}\right)^2+ g^{\theta\theta}\,\left(\frac{dW_\theta}{d\theta}\right)^2=m^2.
\end{equation}
Since the HJ equation can be written as a sum of a function of $r$ only and of $\theta$ only, as
\begin{equation}
\frac{[(r^2+a^2)\,E - a\,L]^2}{\Delta} - m^2\,r^2 - \Delta\,\left(\frac{dW_r}{dr}\right)^2 = \left(a\,E\,\sin\theta - \frac{L}{\sin\theta}\right)^2 + \left(\frac{dW_\theta}{d\theta}\right)^2 + m^2\,a^2\,\cos^2\theta,
\end{equation}
both sides of the equation can be set equal to the same constant $K$,
\begin{equation}
\left(\frac{dW_\theta}{d\theta}\right)^2 = K - \left(a\,E\,\sin\theta - \frac{L}{\sin\theta}\right)^2 - m^2\,a^2\,\cos^2\,\theta,
\end{equation}
and
\begin{equation}
\left(\frac{dW_r}{dr}\right)^2 = \frac{[(r^2+a^2)\,E-a\,L]^2-\Delta\,(K + m^2\,r^2)}{\Delta^2}.
\end{equation}
The motion on a plane with fixed $\theta = \theta_0$ is possible by choosing the angle given by
\begin{equation}
\left(a\,E\,\sin\theta_0 - \frac{L}{\sin\theta_0}\right)^2 + \left(m\,a\,\cos\theta_0\right)^2 = K,
\end{equation}
in which case, the action reads
\begin{equation}
S = E\,t - L\,\phi - \int\,\frac{\sqrt{[(r^2+a^2)\,E-a\,L]^2-\Delta\,(K + m^2\,r^2)}}{\Delta}\,dr.
\end{equation}
According to Eq.~(\ref{taylor_expansion}), expanding the action to its second order in $m^2$ gives
\begin{equation}\label{S1_kerr}
S^{(1)} = \int\,\frac{r^2\,dr}{2\sqrt{[(r^2+a^2)\,E-a\,L]^2-\Delta\,K}},
\end{equation}
and
\begin{equation}\label{S2_kerr}
S^{(2)} = \int\,\frac{\Delta\,r^4\,dr}{4\left\{[(r^2+a^2)\,E-a\,L]^2-\Delta\,K\right\}^{3/2}}.
\end{equation}
When $Q\to 0$, the Kerr-Newman metric reduces to the Kerr metric describing a rotating, charge-less black hole. In this case, the expression for $S^{(1)}$ in Eq.~(\ref{S1_kerr}) reduces to that in Eq.~(14) of Ref.~\cite{wudka01}, where the expression for neutrino oscillations in a Kerr metric is obtained to its first order in $\Delta m^2$.

\subsection{Radial propagation} 

We now deal with the case of radial propagation $L = K = 0$, where we find
\begin{equation} \label{radial_Kerr1}
S^{(1)} = \int\,\frac{r^2\,dr}{2E\,(r^2+a^2)} = \frac{1}{2E}\,\left[r_R-r_A - a\,\left(\arctan\frac{r_B}{a}-\arctan\frac{r_A}{a}\right)\right],
\end{equation}
and
$$S^{(2)} = \int_{r_A}^{r_B}\,\frac{(r^2-r\,r_s+a^2+r_Q^2)\,r^4\,dr}{4E^3\,(r^2+a^2)^3} = $$
\begin{equation} \label{radial_Kerr2}
= \frac{1}{32\,E^3}\,\left[8r + \frac{4 a^2 (r - 2 r_s)-5 \,r_Q^2 r}{a^2 + r^2} + \frac{2a^2\,(r_Q^2 r + a^2\,r_s)}{(a^2 + r^2)^2} - \frac{3 (4 a^2 - r_Q^2) \arctan\frac{r}{a}}{a} - 4 r_s\,\ln\left(a^2 + r^2\right)\right]_{r_A}^{r_B}.
\end{equation}

\subsection{Limits for small and large values of the angular momentum}

Since neither $r_Q$ nor $r_s$ are contained in the expression for the phase difference $\Phi$ at first order in Eq.~\eqref{radial_Kerr1}, we need to discuss limits of this term when $a \ll r_B - r_A$ or $a \gg r_B - r_A$ only.

For $a \ll r_B - r_A$, the phase difference expressed in the first-order term in Eq.~\eqref{radial_Kerr1} reduces to the result obtained in the flat geometry, while in the opposite limit $a \gg r_B - r_A$ we find an additional factor of two. In formulas,
\begin{equation}
S^{(1)} \approx \frac{|r_B - r_A|}{E}\,\begin{cases}
\frac{1}{2},\quad \hbox{for $a \ll |r_B - r_A|$},\\
1,\quad \hbox{for $a \gg |r_B - r_A|$}.
\end{cases}
\end{equation}
The second-order term in Eq.~\eqref{radial_Kerr2} contains all three parameters $r_s$, $r_Q$, and $a$. Here, we notice that for given $r_s$ and $r_Q$ and in the limit $a \ll r_s, r_Q, |r_B-r_A|$, we obtain
\begin{equation}
S^{(2)} \approx  \frac{1}{4E^3}\,\bigg\{\left[\left(r_B - r_A \right)\,\left(1+\frac{r_Q^2}{r_A\,r_B}\right)- r_s\,\ln\frac{r_B}{r_A}\right] + a^2\,\frac{2 r_Q^2 + 4 r^2 - 3 r r_s}{2r^3}\bigg|_{r_A}^{r_B}\bigg\}.
\end{equation}
The term in the squared brackets corresponds to the second-order correction to the neutrino phase difference in the metric of a non-rotating charge black hole.

\subsection{Small values of the angular momentum and the charge}

We now work in the limit $a \sim r_Q \ll r_s \ll |r_B - r_A|$, which corresponds to a Lense-Thirring metric with a slowly-rotating and charged object at the center of the coordinates. Under these conditions, the expressions for the phase shift in Eqs.~\eqref{radial_Kerr1} and~\eqref{radial_Kerr2} reduces to
\begin{equation} \label{expression_small_kerr}
\Phi = \Phi^{\rm Sch} - \frac{\Delta m^2}{2 E}\,\frac{a^2}{r_A\,r_B}\,|r_B-r_A| + \frac{\Delta m^4}{8 E^3}\,\frac{r_Q^2 - 2a^2}{r_A\,r_B}\,|r_B-r_A|,
\end{equation}
where $\Phi^{\rm Sch}$ is the expression for the phase shift in the Schwarzschild metric previously obtained by Godunov and Pastukhov~\cite{godunov09},
\begin{equation}
\Phi^{\rm Sch} = \frac{\Delta m^2}{2E}\,\left|r_B - r_A\right| + \frac{\Delta m^4}{8E^3}\left(r_B - r_A - r_s\,\ln\frac{r_B}{r_A}\right).
\end{equation}
Notice that the gravitational effects due to the mass $r_s$ and the charge $r_Q$ enter the term proportional to $\Delta m^4$ only if the phase oscillation is written in terms of the coordinate difference $r_B - r_A$, while the effects of rotation also enter the first order term $S^{(1)}$.

\subsection{Proper length} 

It is possible to write analytically the expression for the proper length in the Kerr-Newman metric along the plane $\theta = \pi/2$ as
\begin{equation}
L_p = \int_{r_A}^{r_B} \sqrt{-g_{rr}}\,dr' = \sqrt{r^2 - r_s\,r + a^2+r_Q^2}+ \frac{r_s}{2}\,\ln\left(2r-r_s + 2\sqrt{r^2 - r_s\,r + a^2+r_Q^2}\right)\bigg|_{r_A}^{r_B}.\end{equation}
We now work in the limit $a \sim r_Q \ll r_s \ll |r_B - r_A|$, in which
\begin{equation}
L_p = L_p^{\rm Sch} + \frac{a^2+r_Q^2}{r-r_s + \sqrt{r\,(r-r_s)}}\bigg|_{r_A}^{r_B} \approx L_p^{\rm Sch} - \frac{a^2+r_Q^2}{2\,r_A\,r_B}\,|r_B-r_A|.
\end{equation}
Here, we introduced the proper length in the Schwarzschild metric
\begin{equation}
L_p^{\rm Sch} = \sqrt{r^2 - r_s\,r} + \frac{r_s}{2}\,\ln\left(2r-r_s + 2\sqrt{r^2-r_s\,r}\right)\bigg|_{r_A}^{r_B} \approx r_B-r_A + \frac{r_s}{2}\,\ln\frac{r_B}{r_A}.
\end{equation}
The coordinate length is then written in terms of the proper length as
\begin{equation}
|r_B-r_A| \approx L_p\,\left(1 + \frac{a^2+r_Q^2}{2\,r_A\,r_B}\right) -  \frac{r_s}{2}\,\ln\frac{r_B}{r_A}.
\end{equation}
Following Fornengo {\it et al.}~\cite{fornengo97}~\cite{fornengo99} and Crocker {\it et al.}~\cite{crocker04}, we introduce the local energy $E^{\rm loc}(r_B)$ measured by an observer at rest at the position $r_B$, as 
\begin{equation}
E^{\rm loc}(r_B) = E/\sqrt{g_{tt}(r_B)} = \frac{r_B\,E}{\sqrt{r_B^2- r_s\,r_B + r_Q^2}}\approx E \,\left(1+\frac{r_s}{2r_B} - \frac{r_Q^2}{2r_B^2}\right).
\end{equation}
At first-order, the phase shift in Eq.~\eqref{expression_small_kerr} in the local coordinates introduced is then written as
\begin{equation} \label{expression_small_kerr1}
\Phi = \frac{\Delta m^2\,L_p}{2E^{\rm loc}(r_B)}\,\left[1 + \frac{r_Q^2\,L_p}{2r_A\,r_B^2} - \frac{a^2}{2\,r_A\,r_B} -  \frac{r_s}{2}\,\left(\frac{1}{L_p}\ln\frac{r_B}{r_A} + \frac{1}{r_B}\right)\right],
\end{equation}
and the expression reduces to the ordinary result $\Phi = \Delta m^2\,L_p/2E^{\rm loc}(r_B)$ in the Schwarzschild limit $r_s = a = r_Q = 0$.

\section{Application to the Friedmann-Robertson-Walker metric} \label{sec_FRW}

\begin{figure}[h!]
\begin{center}
  \includegraphics[width=10cm]{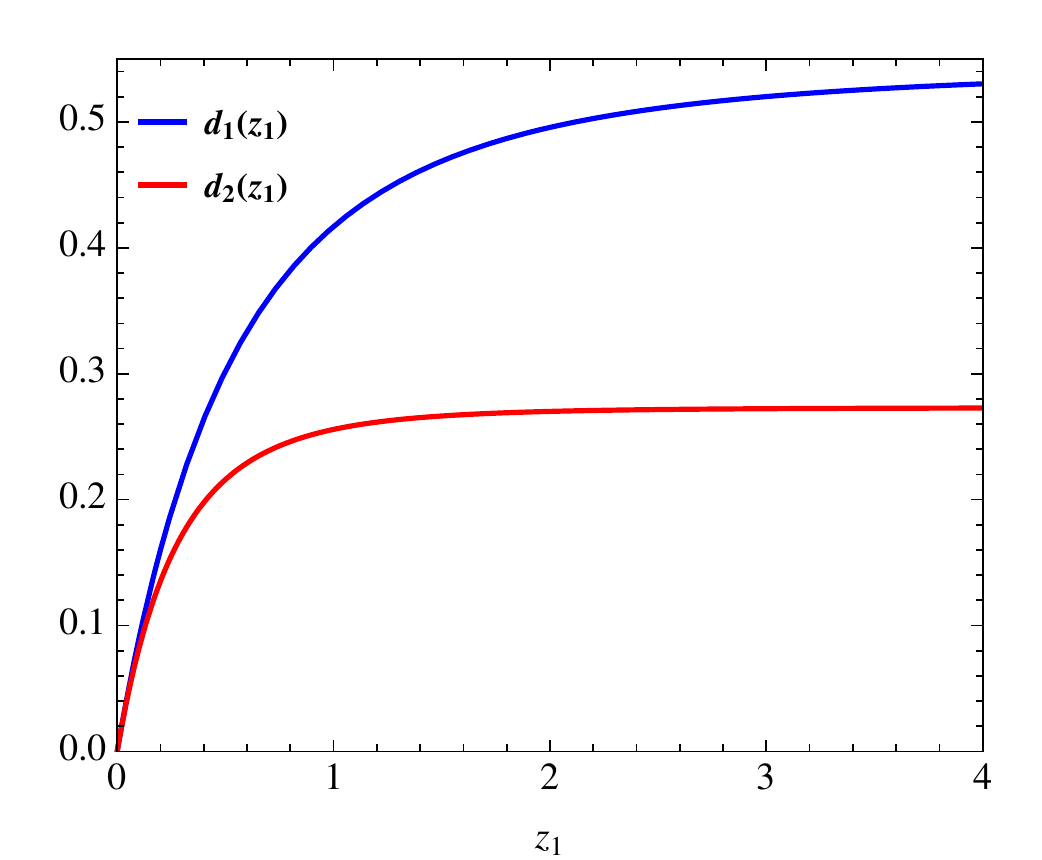}
\caption{The functions $d_1(z_1)$ (blue solid line) and $d_2(z_1)$ (red solid line), in units of $1/H_0$, computed from Eq.~(\ref{distance1}) and Eq.~(\ref{distance2}), respectively. }
\label{plot_distance}
\end{center}
\end{figure}

We consider the line element for a flat, conformal FRW metric
\begin{equation}
ds^2 = a^2(\eta)\,(d\eta^2 - dr^2- r^2d\theta^2 - r^2\sin^2\theta\,d\phi^2),
\end{equation}
where $\eta$ is the conformal time. The scale factor $a = a(\eta)$ describes the expansion of the Universe through the Friedmann equation, see Eq.~(\ref{frieldmann_eq}) below, and it is normalized so that at present time it is $a(0) = 1$. Azymuthal angular momentum $L$ is a conserved quantity in the FRW metric, while the energy is not since $\partial_\eta$ is not a Killing vector of the FRW metric. We look for the additional invariants in the metric by writing the neutrino action as
\begin{equation}
S = W(\eta) - W_r(r) - W_\theta(\theta)  - L\,\phi,
\end{equation}
and the HJ Eq.~(\ref{HJ}) is
\begin{equation}
\left(\frac{dW_\eta}{d\eta}\right)^2 -\left(\frac{dW_r}{dr}\right)^2 - \frac{1}{r^2}\,\left(\frac{dW_\theta}{d\theta}\right)^2 - \frac{L^2}{r^2\,\sin^2\theta} = m^2\,a^2(\eta).
\end{equation}
This differential equation is separable into a set of three equations as
$$\left(\frac{dW_\eta}{d\eta}\right)^2 = m^2\,a^2(\eta) + E^2,$$

$$\left(\frac{dW_r}{dr}\right)^2 = E^2 - \frac{K^2}{r^2},$$

$$\left(\frac{dW_\theta}{d\theta}\right)^2 = K^2 - \frac{L^2}{\sin^2\theta},$$

where we introduced two new constants $E$ and $K$. Fixing the plane in which the motion occurs by setting $\sin\theta = L/K$, the action is
\begin{equation}\label{action_FRW}
S = \int\,\sqrt{m^2\,a^2(\eta) + E^2}\,d\eta - \int\,\sqrt{E^2 - \frac{K^2}{r^2}}\,dr - L\,\phi.
\end{equation}
Eq.~(\ref{action_FRW}), expanded to the second order in $m^2$ according to Eq.~(\ref{taylor_expansion}), gives
\begin{equation}\label{FRW_expansion}
S^{(1)} = \int \frac{a^2\,d\eta}{2E},\quad\hbox{and}\quad S^{(2)} = \int \frac{a^4\,d\eta}{4E^3}.
\end{equation}
To simplify these integrals, we use the definition for the red-shift $z$ in terms of the conformal factor,
\begin{equation}
1+z = \frac{1}{a(\eta)},
\end{equation}
together with the Friedmann equation
\begin{equation} \label{frieldmann_eq}
H(z) = H_0\, f(z),
\end{equation}
where the Hubble rate is
\begin{equation} \label{friedmann}
H(z) = \frac{1}{a^2(\eta)}\,\frac{d\,a(\eta)}{d\eta} = -\frac{dz}{d\eta}.
\end{equation}
Here, $H_0 = H(z=0)$ is the present Hubble rate, and we defined the function
\begin{equation}
f(z) = \sqrt{\Omega_r\,(1+z)^4+\Omega_m\,(1+z)^3+\Omega_k\,(1+z)^2+\Omega_\Lambda},
\end{equation}
which accounts for the content of the Universe and the equations of state for radiation, cold matter, curvature, and dark energy, respectively. With these substitutions, the relative phase in Eq.~(\ref{relative_phase_expanded}) for two neutrino generations is
\begin{equation}\label{FRW_phase}
\Phi = \frac{\Delta m^2}{2E}\,d_1(z_1) + \frac{\Delta m^4}{8E^3}\,d_2(z_1),
\end{equation}
where $d_1(z_1)$ is the cosmological distance from the source, at redshift $z_1$, to the detector at $z=0$,
\begin{equation} \label{distance1}
d_1(z_1)  = \frac{1}{H_0}\int_0^{z_1}\frac{dz}{(1+z)^2f(z)};
\end{equation}
we also defined the function
\begin{equation} \label{distance2}
d_2(z_1)  = \frac{1}{H_0}\,\int^{z_1}_0\frac{dz}{(1+z)^4\,f(z)}.
\end{equation}
For close ($z_1 \ll 1$) astrophysical sources of neutrinos, we recover the usual formula
\begin{equation}
d_1(z_1) \approx d_2(z_1) \approx \frac{z_1}{H_0} = r_B-r_A.
\end{equation}
For more distant sources the curvature of the Universe affects measurements, as shown in Fig.~\ref{plot_distance} where we neglect the curvature and radiation terms $\Omega_k = \Omega_r = 0$, and we use $\Omega_m = 0.27$ and $\Omega_\Lambda = 1 - \Omega_m$. The plot in Fig.~\ref{plot_distance} shows that both $d_1(z_1)$ and $d_2(z_1)$ reach a constant value, thus the phase difference results constant for sources with redshift $z_1 > 2$. The constancy of the function $d_1(z_1)$ for large values of $z_1$ has also been noticed by Silk and Stodolsky~\cite{silk2006}, in the context of the dilution of the cosmological flux of weakly interacting particles.

\begin{figure}[h!]
\begin{center}
  \includegraphics[width=10cm]{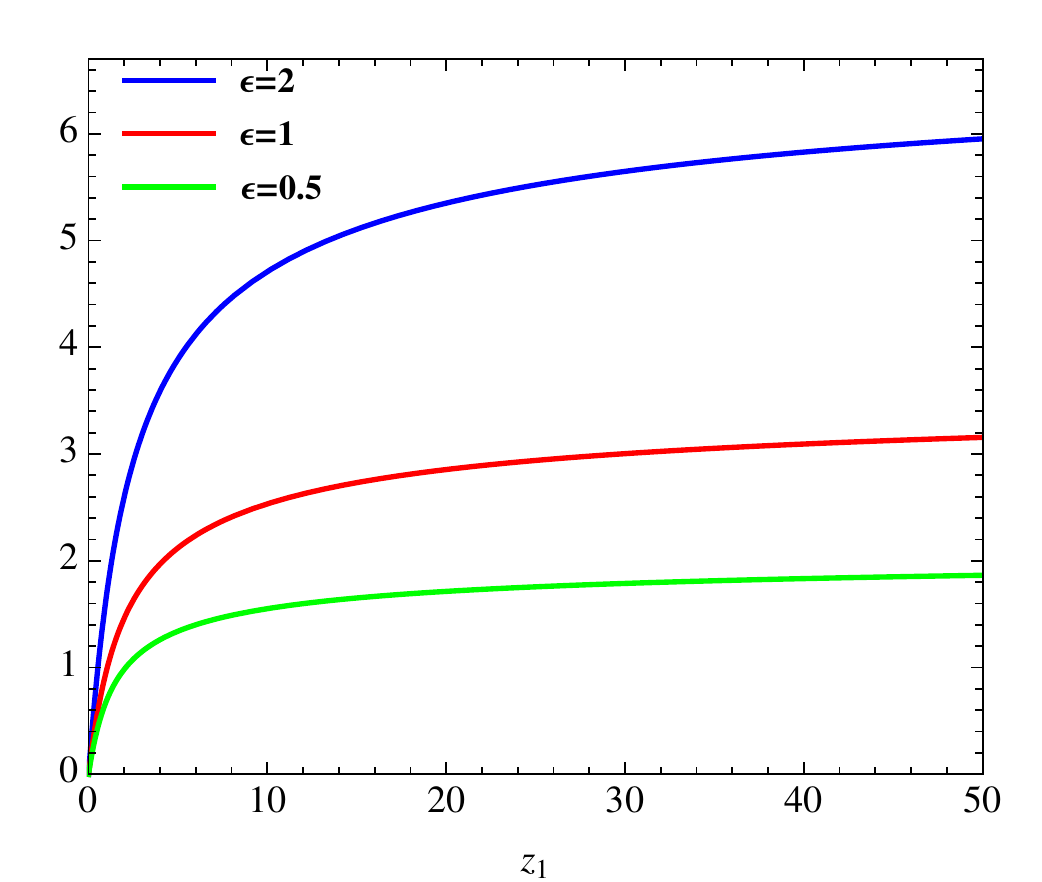}
\caption{The function $d(z_1)$, in units of $1/H_0$, computed from Eq.~(\ref{distance_full}) for $\epsilon = 0.5$ (green solid line), $\epsilon = 1$ (red solid line), and $\epsilon = 2$ (blue solid line). }
\label{plot_distance_full}
\end{center}
\end{figure}

This behavior is retained even if we restrain from expanding the action in Eq.~(\ref{action_FRW}) in Taylor series. In fact, writing the action as
\begin{equation}
S = m\,d(z_1) - \int\,\sqrt{E^2 - \frac{K^2}{r^2}}\,dr - L\,\phi,
\end{equation}
where, setting $\epsilon = E/m$, we defined
\begin{equation} \label{distance_full}
d(z_1) = \frac{1}{H_0}\,\int_0^{z_1}\,\sqrt{\frac{1}{(1+z)^2} + \epsilon^2}\,\frac{dz}{f(z)},
\end{equation}
we obtain that the function $d(z_1)$ reaches a constant value when $z_1 \gg 1$, as shown in Fig.~\ref{plot_distance_full} for different values of $\epsilon$.

To the best of our knowledge, the expression for the phase difference in the oscillation of neutrino mass eigenstates over a FRW metric in Eq.~(\ref{FRW_phase}) has never been derived before.

\section{Concluding remarks}

In this paper we have reviewed the mathematics of the Dirac equation in a curved space-time and its application to neutrino oscillations. In particular, we have derived the method of calculating the phase shift in flavor neutrino oscillations by a Taylor-expansion of the action in orders of $m^2$.

In Sec.~\ref{sec_kerr}, we have applied this method by evaluating the correction to the phase difference of neutrino mass eigenstates due to the gravitational field produced by a rotating and charged black hole, described by the Kerr-Newman metric. We have shown that, for the case of radial propagation, the effects of the black hole rotation are present at the first order term in the mass splitting $\Delta m^2$ and dominate the phase difference with respect to the charge, which appear at the second order. For a charge-less and rotation-less black hole, we have recovered the results obtained in the previous literature.

In addition, in Sec.~\ref{sec_FRW} we have applied the Taylor series method to the case of cosmological neutrinos propagating in an expanding, flat Friedmann-Robertson-Walker metric. After discussing the equations for the invariants of motion, we have obtained an expression for the phase difference for neutrino mass eigenstates in a $\Lambda$-CDM model. We have shown that, for distant sources $z_1 \gg 1$, the phase difference is independent of distance, thus neutrino oscillations cannot be used to infer the nature of the beam. For close sources $z_1 \ll 1$, we have recovered the usual result that the phase difference is proportional to the distance itself.

\section*{Acknowledgments}
The author would like to thank Paolo Gondolo (U. of Utah) for useful discussions on neutrino oscillations.

\end{document}